\documentclass[journal=jpccck,manuscript=article,layout=twocolumn]{achemso}
\usepackage{graphicx}
\usepackage{epstopdf}
\usepackage{amssymb,amsmath}
\usepackage{siunitx}
\usepackage{threeparttable}
\usepackage[version=3]{mhchem}
\usepackage{xfrac}
\usepackage{footmisc}

\author{Mie Andersen}
\email{mie.andersen@ch.tum.de}
\affiliation{Chair for Theoretical Chemistry and Catalysis Research Center, Technische Universit{\"a}t M{\"u}nchen, Lichtenbergstr. 4, 85747 Garching, Germany}
\author{Juan Santiago Cingolani}
\affiliation{Chair for Theoretical Chemistry and Catalysis Research Center, Technische Universit{\"a}t M{\"u}nchen, Lichtenbergstr. 4, 85747 Garching, Germany}
\author{Karsten Reuter}
\affiliation{Chair for Theoretical Chemistry and Catalysis Research Center, Technische Universit{\"a}t M{\"u}nchen, Lichtenbergstr. 4, 85747 Garching, Germany}

\title{Ab initio thermodynamics of hydrocarbons relevant to graphene growth at solid and liquid Cu surfaces}

\begin{document}

\begin{tocentry}
\includegraphics[width=\textwidth]{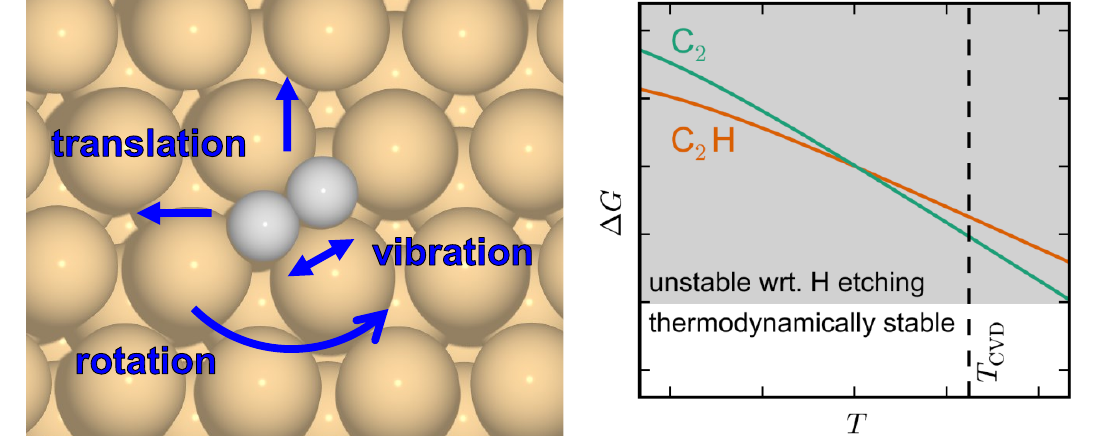}
\end{tocentry}

\begin{abstract}
Using \textit{ab initio} thermodynamics, the stability of a wide range of hydrocarbon adsorbates under various chemical vapor deposition (CVD) conditions (temperature, methane and hydrogen pressures) used in experimental graphene growth protocols at solid and liquid Cu surfaces has been explored. At the employed high growth temperatures around the melting point of Cu, we find that commonly used thermodynamic models such as the harmonic oscillator model may no longer be accurate. Instead, we account for the translational and rotational mobility of adsorbates using a recently developed hindered translator and rotator model or a two-dimensional ideal gas model. The thermodynamic considerations turn out to be crucial for explaining experimental results and allow us to improve and extend the findings of earlier theoretical studies regarding the role of hydrogen and hydrocarbon species in CVD. In particular, we find that smaller hydrocarbons will completely dehydrogenate under most CVD conditions. For larger clusters our results show that metal-terminated and hydrogen-terminated edges have very similar stabilities. While both cluster types might thus form during the experiment, we show that the low binding strength of clusters with hydrogen-terminated edges could result in instability towards desorption.
\end{abstract}

%\noindent
%\textbf{Keywords:} graphene, chemical vapor deposition, ab initio thermodynamics, density functional theory, liquid metal catalysis 

\section{Introduction}
Since the first experimental synthesis and characterization of graphene in 2004 \cite{Novoselov2004}, this two-dimensional (2D) material has attracted much interest owing to its remarkable structural and electronic properties such as its strength and flexibility and the exceptionally high electrical conductivity \cite{Novoselov2005,Allen2010}. However, a full exploitation of these properties requires the development of effective mass production techniques. CVD on solid metal surfaces, in particular Cu foils and thin films, has been established as an important and efficient synthesis method, but often the samples produced suffer from defects and impurities \cite{Li2009,Bhaviripudi2010,Vlassiouk2011,Kim2012,Li2016,Huet2017}. Recent experimental evidence suggests that using a liquid Cu surface instead of a solid one enables the fast growth of large single-crystalline and single-layered graphene flakes of very high structural quality \cite{Geng2012,Wu2012,Wu2013,Geng2014,Zeng2016,Xue2019,Zheng2019}. While these findings are very promising for industrial-scale production of high-quality graphene samples, the reason for the improved catalytic properties of liquid Cu, and the catalytic properties of liquid metals in general, is still poorly understood.

In a recent study molecular dynamics (MD) simulations based on density-functional tight-binding (DFTB) suggested that the high structural quality of graphene grown on liquid Cu could be related to defect-healing mechanisms that are only possible in the liquid state \cite{Li2014}. However, the high computational demands of MD simulations severely limits the timescales that can be reached in the simulation and restricts the system complexity that can be taken into account, even when employing cheaper semi-empirical methods such as DFTB. In particular, this study neglected the role of typical gas-phase reactants such as methane and hydrogen and considered only growth initiated by deposition of C$_2$ dimers onto the surface. Other studies based on density-functional theory (DFT) calculations have considered graphene growth on solid Cu facets and taken into account also reaction steps related to methane decomposition and the role of the typically large concentration of hydrogen present in experimental growth setups \cite{Zhang2011,Zhang2014,Shu2015,Li2017}.

An efficient consideration of such gas-phase pressures and finite temperatures can also be achieved by combining the \textit{ab initio} calculations with thermodynamic considerations to evaluate the Gibbs free energies of reaction intermediates under varying reaction conditions, also known as \textit{ab initio} thermodynamics \cite{Reuter2005}. In contrast to MD simulations, this represents the limit of infinite timescales where a constrained thermodynamic equilibrium between the gas-phase reactants and the considered surface adsorbate or adsorbate configuration has been reached, while disregarding further chemical reactions between the gas-phase or surface species. Compared to typical catalytic systems investigated previously with this methodology \cite{Reuter2003,Monder2010,Exner2014}, a particularity of graphene growth on Cu is that diffusion and rotation barriers of hydrocarbon species are typically rather low, while the temperatures employed are very high, e.g.\ around 1300$-$1400 K. This necessitates additional care in the evaluation of free energy contributions, since often used models such as treating the adsorbate as an immobile harmonic oscillator (HO) might no longer be valid.

In this work we improve and extend previous \textit{ab initio} thermodynamics studies of hydrocarbon adsorbates under typical graphene growth conditions employed in CVD on solid Cu surfaces and additionally consider also the case of liquid Cu CVD. We find that the thermodynamic considerations are critically important to explain experimental findings regarding e.g.\ effects of varying the methane to hydrogen ratio in the reactant mixture, susceptibility to multi-layer growth, activation energy for graphene growth and the high structural quality of graphene flakes grown on liquid Cu. To this end, theoretical models in the literature to date can have led to wrong conclusions due to either the complete negligence of thermodynamic considerations or the use of simplifying HO models.

\section{Computational details}

\subsection{Density-functional theory}
\begin{figure}
\centering
\includegraphics[width=\columnwidth]{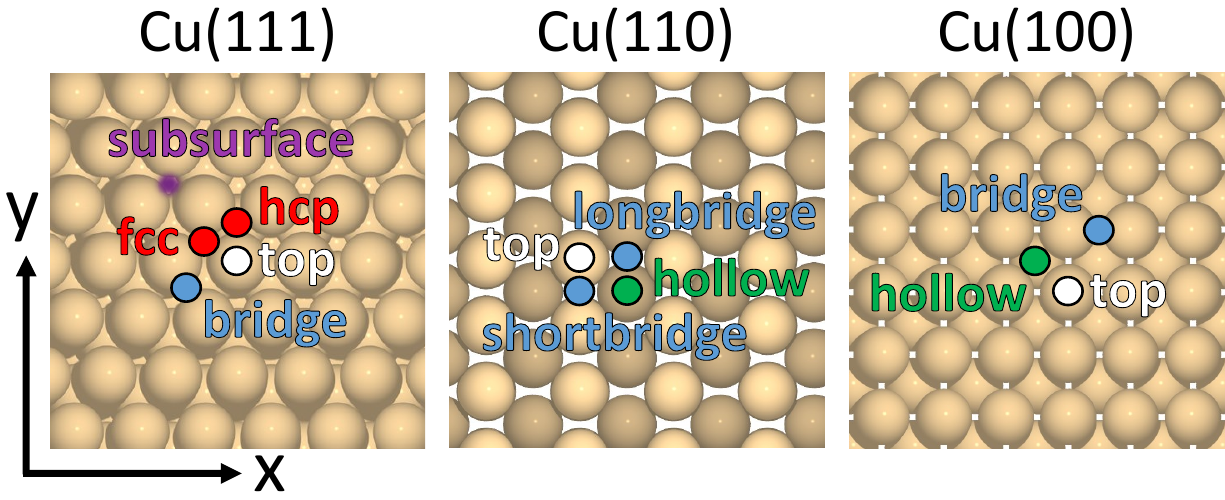}
\caption{Illustration of the three considered Cu facets as well as their high-symmetry adsorption sites.}
\label{Fig:1}
\end{figure}

The DFT calculations were carried out with the plane-wave code Quantum ESPRESSO \cite{Giannozzi2017} v.6.3 using ultrasoft pseudopotentials\footnote{Ultrasoft pseudopotentials were taken from the Quantum ESPRESSO pseudopotential library and were generated using the ``atomic'' code by A. Dal Corso in 2012 (v.5.0.2 svn rev. 9415)}. Exchange and correlation effects were described using the PBE functional \cite{Perdew1996}, while van der Waals (vdW) interactions were accounted for using the semi-empirical D3 correction scheme by Grimme \cite{Grimme2010}.
For the three considered low-index facets Cu(111), Cu(110), and Cu(100) and the hydrocarbon clusters containing up to 6 C atoms the lateral size of the supercell was chosen large enough that the clusters in periodic images were separated by at least 10 $\text{\AA}$. This separation was reduced for the larger hydrocarbons to keep the computational cost tractable. The smallest separations used were 8.3 $\text{\AA}$ for C$_{24}$ and about 6.4 $\text{\AA}$ for the hydrogen-terminated C$_{24}$H$_{12}$ cluster and the hydrogen-terminated graphene edge (see the Supporting Information (SI) for a detailed overview of all structures). The PBE-optimized Cu bulk lattice constant of 3.648 $\text{\AA}$ was used throughout. Graphene on Cu(111) was modelled in a $(1 \times 1)$ cell, where the lateral size of the cell was fixed to the PBE-optimized Cu nearest neighbor distance of 2.580 $\text{\AA}$ and the graphene lattice constant was adjusted accordingly. The number of metal layers were four (Cu(111)), eight (Cu(110)), and five (Cu(100)), with adsorption on one side of the slab. In all cases the lower two metal layers were kept fixed in their bulk-truncated positions, while the upper layers and the hydrocarbon cluster were relaxed until the maximum force on each atom fell below 0.01\,eV/$\text{\AA}$. A vacuum region of at least 16 $\text{\AA}$ perpendicular to the surface separated the slab from its periodic images and a dipole correction was applied \cite{Bengtsson1999}. The Brillouin zone was sampled with a $(n \times m)$ grid of \textbf{k}-points, where along each ($x$,$y$) cell direction at least 31/$a$ \textbf{k}-points were used, $a$ being the cell length in $\text{\AA}$. Cutoffs of 500\,eV and 5000\,eV were used for the orbitals and the charge density, respectively.
For each hydrocarbon cluster containing up to 6 C atoms several different high-symmetry adsorption sites (see Fig.\ \ref{Fig:1}) were tested along with various rotations within the site, if applicable, in order to find the most stable adsorption configuration. For the larger hydrocarbon clusters C$_{13}$, C$_{21}$, and C$_{24}$ the adsorption configuration was taken from Ref.\ \cite{Zhong2016}, while for the graphene edges the structures used were taken from Ref.\ \cite{Zhang2014}. The graphene edges were optimized with a cell length of one graphene lattice constant along the periodic zigzag direction, while for the calculation of the vibrational frequencies the cell was repeated four times along the periodic direction, as was also done in Ref.\ \cite{Zhang2014}. Energies and vibrational frequencies of gas-phase radical species were calculated spin-polarized, all other calculations were conducted non-spin-polarized.

%%%% FIGURE %%%% 
%%%% FIGURE %%%% 

Transition state (TS) energies were calculated using the climbing-image nudged-elastic-band (CI-NEB) method \cite{Henkelman2000} and using a convergence threshold of 0.05\,eV/$\text{\AA}$ for the optimization. The harmonic vibrational frequencies of the hydrocarbon clusters and graphene edges were calculated using the Atomic Simulation Environment (ASE) code v.3.13 \cite{Hjorth_Larsen_2017}, and the phonon density of states of graphene was calculated using the PHonon package from Quantum ESPRESSO, the PBE-optimized graphene lattice constant of 2.463 $\text{\AA}$ and a $(32 \times 32)$ \textbf{k}-point grid. All energies and vibrational frequencies are listed in Sec.\ S1 of the SI.

\subsection{Ab initio thermodynamics}
The Gibbs free energy of formation $\Delta G_{\rm f}$ of a hydrocarbon species adsorbed to a Cu surface is evaluated using the following formula
\begin{equation}\label{eq:1}
\begin{split}
\Delta G_{\rm f}(T,p_{\rm CH_4},p_{\rm H_2}) = &\ \frac{G_{\rm Cu+ads}(T) - G_{\rm Cu}(T)}{x} \\
- &\ G_{\rm CH_4}(T,p_{\rm CH_4}) \\
- &\ \left( \frac{y}{2 x}-2 \right) G_{\rm H_2}(T,p_{\rm H_2}) \quad . \\
\end{split}
\end{equation}
Here $T$ is the temperature, $p_{\rm CH_4}$ and $p_{\rm H_2}$ are the partial pressures of the gas-phase reactants methane and hydrogen, respectively, $G_{\rm Cu+ads}$ is the Gibbs free energy of the adsorbate on the Cu surface, $x$ and $y$ are the number of C and H atoms in the adsorbate, respectively, $G_{\rm Cu}$ is the Gibbs free energy of the Cu surface, and $G_{\rm CH_4}$ and $G_{\rm H_2}$ are the Gibbs free energies of a methane and a hydrogen molecule in the gas phase, respectively. The latter two were evaluated in the ideal gas approximation as implemented in the IdealGasThermo class in the ASE thermochemistry module using the DFT-calculated energy along with experimental vibrational frequencies \cite{NIST}. Internal energy and entropy corrections to the Cu surface in $G_{\rm Cu+ads}$ and $G_{\rm Cu}$ are not considered, as these are assumed to be approximately equal with and without the adsorbate and therefore cancel each other out to a large degree in Eq.\ \ref{eq:1}. With the current definition and considering Gibbs free energies to be given as negative numbers (more negative implying higher stability), a $\Delta G_{\rm f} > 0$ indicates a thermodynamic instability of an adsorbed hydrocarbon cluster with respect to decomposition into gas-phase methane and hydrogen. Since the investigated clusters all contain less hydrogen per carbon atom than a methane molecule, this thermodynamic instability means in practice that the cluster is unstable with respect to the reaction with hydrogen to form methane at the surface, which then desorbs to the gas phase. In accordance with experimental literature, we will refer to this process as hydrogen etching (see also Scheme \ref{Scheme1} below and accompanying discussion).

\begin{figure*}
\centering
\includegraphics[width=\textwidth]{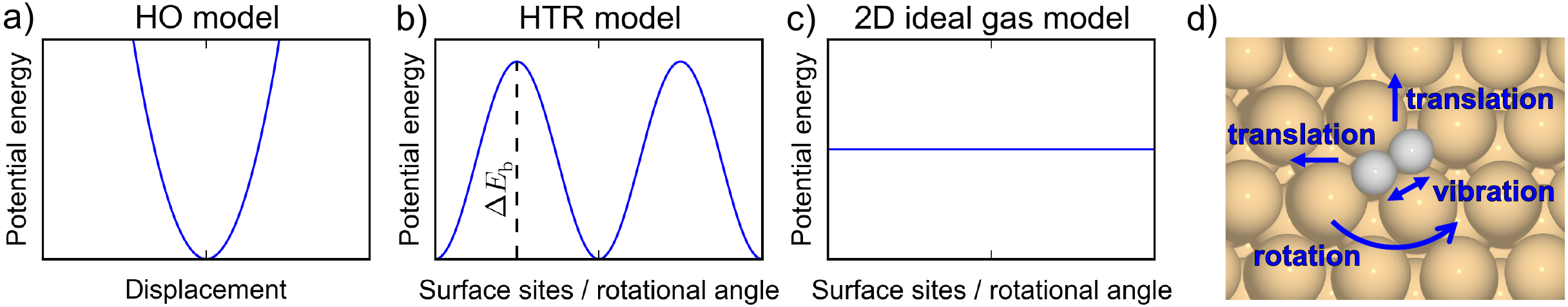}
\caption{Illustration of the assumed shape of the potential energy surface in a) the harmonic oscillator (HO) model, b) the hindered translator / rotator (HTR) model with diffusion / rotation barrier $\Delta E_{\rm b}$, and c) the 2D ideal gas model where $\Delta E_{\rm b}$ equals zero. d) Illustration of the two translational, one rotational, and one of the three vibrational degrees of freedom for C$_2$ at Cu(111).}
\label{Fig:2}
\end{figure*}

For the evaluation of the Gibbs free energy of the adsorbates we consider various models from the literature. The approach taken in most literature studies is to evaluate adsorbate free energies in the harmonic oscillator (HO) model (see Fig.\ \ref{Fig:2}(a)). In this model the adsorbate is assumed to remain immobile at the most stable adsorption site, and all degrees of freedom are treated as harmonic vibrations. For this we used the HarmonicThermo class from the ASE thermochemistry module with DFT-calculated energies and frequencies. 

%%%% FIGURE %%%% 
%%%% FIGURE %%%% 

However, the assumption of immobile adsorbates breaks down for low diffusion or rotation barriers and high temperatures. In this case it has recently been shown by Campbell and coworkers that the free energy may be more accurately evaluated by treating the adsorbate as a hindered translator / rotator (HTR) \cite{Sprowl2016} (see Fig.\ \ref{Fig:2}(b)). In the HTR model the adsorbate is assumed to have two hindered translational modes parallel to the surface, one hindered rotational mode around an axis perpendicular to the surface (except for atoms and linear molecules adsorbed perpendicular to the surface), as well as the remaining $3N-m$ harmonic vibrational modes (see Fig.\ \ref{Fig:2}(d)), where $N$ is the number of atoms in the adsorbate and $m$ is two or three depending on whether only the two translational modes or also the additional rotational mode exist. In practice this is carried out by replacing the lowest $m$ calculated frequencies with the corresponding translational and rotational modes. We used the HTR model implemented in the HinderedThermo class in the ASE thermochemistry module with a few modifications: i) we included a user input to indicate whether a rotational mode should be included for the adsorbate or not, ii) we allowed for diffusion barriers and distances between diffusional minima to be different along the ($x$,$y$) cell directions (important for Cu(110), see Fig.\ \ref{Fig:1}), and iii) we removed the concentration-related entropy calculated for the standard state surface concentration proposed by Campbell and coworkers \cite{Campbell2016}, since we consider here only the stability of individual adsorbate species at the Cu surface. The inputs to the HTR model are the DFT-calculated energies, frequencies, diffusion barriers, distances between diffusional minima, rotation barriers and number of rotational minima (see Table \ref{tab:1}). 

%%%% TABLE %%%% 
\begin{table*}
  \centering
  \begin{threeparttable}
  \begin{tabular}{lccccccr}
	  \hline\hline
		Facet & Species & Site & $\Delta E_{\rm f}$ & $\Delta E_{\rm diff}$ & $a_{\rm diff}$ & $\Delta E_{\rm rot}$ & $n_{\rm rot}$ \\
		\hline
		(111) & C & subsurface & 3.51 & 0.51 & 2.58 & & \\
		& CH & fcc & 2.47 & 0.12 & 1.49 & & \\
		& C$_2$ & bridge & 2.50 & 0.56\tnote{a} & 1.49 & 0.56\tnote{a} & 6 \\
		& C$_2$H & hcp & 2.08 & 0.30\tnote{a} & 1.49 & 0.30\tnote{a} & 6 \\
		\hline
		(110) & C & hollow & 3.40 & 0.89, 0.37 & 1.82, 1.29 & & \\
		& CH & hollow & 2.35 & 0.96, 0.21 & 3.65, 1.29 & & \\
		& C$_2$ & hollow & 2.20 & 1.88, 0.70 & 3.65, 2.58 & 0.65 & 2 \\
		& C$_2$H & hollow & 1.96 & 0.87, 0.25 & 3.65, 2.58 & 0.35 & 2 \\
		\hline
		(100) & C & hollow & 2.89 & 1.46 & 2.58 & & \\
		& CH & hollow & 1.92 & 1.27 & 2.58 & & \\
		& C$_2$ & hollow & 2.39 & 0.78 & 2.58 & 0.29 & 4 \\
		& C$_2$H & hollow & 1.94 & 0.54 & 2.58 & 0.09 & 4 \\
		\hline\hline
  \end{tabular}
  \begin{tablenotes}
  \item [a] For C$_2$ and C$_2$H at Cu(111) the transition state for diffusion and rotation is the same.
  \end{tablenotes}
  \end{threeparttable}
  \caption{Most stable adsorption site, formation energy ($\Delta E_{\rm f}$), diffusion barrier ($\Delta E_{\rm diff}$), distance between diffusional minima ($a_{\rm diff}$), rotation barrier ($\Delta E_{\rm rot}$) and number of rotational minima ($n_{\rm rot}$) for hydrocarbon species at the three low-index Cu facets. For the Cu(110) facet the diffusion barriers and distances are given separately along the (x,y) directions, respectively (see Fig.\ \ref{Fig:1}). Energies are in eV and distances are in $\text{\AA}$.}
  \label{tab:1}
\end{table*}
%%%% TABLE %%%% 

In the case where all diffusion and rotation barriers go to zero, the HTR model reduces to the 2D ideal gas model \cite{Sprowl2016} (see Fig.\ \ref{Fig:2}(c)). The evaluation of rotational and translational entropy in the 2D ideal gas model requires as input the number of rotational minima and the distance between diffusional minima, respectively. When applying this model to a liquid Cu surface we set the number of rotational minima to one and the distance between diffusional minima to the optimized Cu fcc nearest neighbor distance (2.580 $\text{\AA}$).

The Gibbs free energy of graphene was evaluated using the CrystalThermo class in the ASE thermochemistry module based on the DFT-calculated phonon density of states (see Fig.\ S1 in the SI). To model graphene adsorbed on Cu(111) we added to this free energy the DFT-calculated graphene adsorption energy (see Table \ref{tab:2}).

DFT formation energies $\Delta E_{\rm f}$ quoted in the text are calculated as
\begin{equation}\label{eq:2}
\Delta E_{\rm f} = \frac{E_{\rm Cu+ads} - E_{\rm Cu}}{x}- E_{\rm CH_4} - \left( \frac{y}{2 x}-2 \right) E_{\rm H_2}
\end{equation}
and DFT adsorption energies $\Delta E_{\rm ads}$ are calculated as
\begin{equation}\label{eq:3}
\Delta E_{\rm ads} = E_{\rm Cu+ads} - E_{\rm Cu} - E_{\rm ads} \quad ,
\end{equation}
where $E_{\rm Cu+ads}$ is the energy of the adsorbate on the Cu surface, $E_{\rm CH_4}$ and $E_{\rm H_2}$ are the energies of methane and hydrogen in the gas phase, respectively, and $E_{\rm ads}$ is the energy of the adsorbate in the gas phase.

The desorption rate constant $k_{\rm des}$ of an adsorbate in the 2D ideal gas model is calculated using the derivation based on transition state theory (TST) from Ref.\ \cite{Campbell2016}. It is assumed that the activation energy for adsorption is negligible and that the sticking probability is unity, in which case the TS is the molecule with its center of mass constricted to lie in a plane parallel to the surface at a distance sufficiently far from the surface that the surface-molecule interaction is negligible at any rotational angle or translation. In TST the partition function for translational motion of the TS along the reaction coordinate should be removed as it is already accounted for. The partition function for translational motion parallel to the surface is equal for the TS and the adsorbate in the 2D ideal gas model and therefore cancels out. This leaves the following expression for the desorption rate constant
\begin{equation}\label{eq:4}
k_{\rm des} = \frac{k_{\rm B} T}{h} \frac{1}{q^0_{\rm ads,vib,z}} \frac{q^0_{\rm TS,int}}{q^0_{\rm ads,int}} \exp \left( \frac{\Delta E^0_{\rm ads}}{k_{\rm B} T} \right) \quad ,
\end{equation}
where $k_{\rm B}$ is the Boltzmann constant, $h$ is the Planck constant, $q^0_{\rm ads,vib,z}$ is the partition function for vibrational motion of the adsorbate perpendicular to the surface, $q^0_{\rm TS,int}$ and $q^0_{\rm ads,int}$ are the partition functions for the internal degrees of freedom (rotations, vibrations and electronic excitations in case of gas-phase radicals with non-zero spin) of the TS and adsorbate, respectively, and $\Delta E^0_{\rm ads}$ is the adsorption energy of the adsorbate. The 0 in the partition functions signifies that they are evaluated relative to their zero-point energies and the 0 in the adsorption energy signifies that the zero-point energy-corrected value is used. The lifetime $\tau$ of an adsorbate at the Cu surface is then estimated as $1/k_{\rm des}$.

\section{Results and discussion}
\subsection{Method assessment}
We begin by assessing the reliability of our computational setup for describing vdW interactions between Cu(111) and graphene. Table \ref{tab:2} compares our work using Quantum ESPRESSO and PBE with D3 vdW corrections to previous studies from the literature using the GPAW code with exact exchange and the random phase approximation (EXX+RPA) or the parametrized meta-generalized gradient approximation M06-L functional. It is seen that the agreement of both the adsorption energy and the adsorption distance to the literature studies is excellent, making us confident that we reliably describe the interaction of graphitic adsorbates with Cu surfaces.

%%%% TABLE %%%% 
\begin{table}
  \centering
  \begin{threeparttable}
  \begin{tabular}{lcc}
	  \hline\hline
		Code / functional & $\Delta E_{\rm ads}$ & $d$ \\
		\hline
		Quantum ESPRESSO / PBE+D3 & -60 & 3.22 \\
		GPAW / EXX+RPA\tnote{a} & -62 & 3.25 \\
		GPAW / M06-L\tnote{b} & -61 & 3.32 \\
		\hline\hline
  \end{tabular}
  \begin{tablenotes}
  \item [a] From Ref.\ \cite{Olsen2011}.
  \item [b] From Ref.\ \cite{Andersen2012}.
  \end{tablenotes}
  \end{threeparttable}
  \caption{Adsorption energy ($\Delta E_{\rm ads}$ in meV / C atom) and adsorption distance ($d$ in $\text{\AA}$) for graphene on Cu(111) calculated using various codes and functionals.}
  \label{tab:2}
\end{table}
%%%% TABLE %%%% 

\begin{figure}
\centering
\includegraphics[width=\columnwidth]{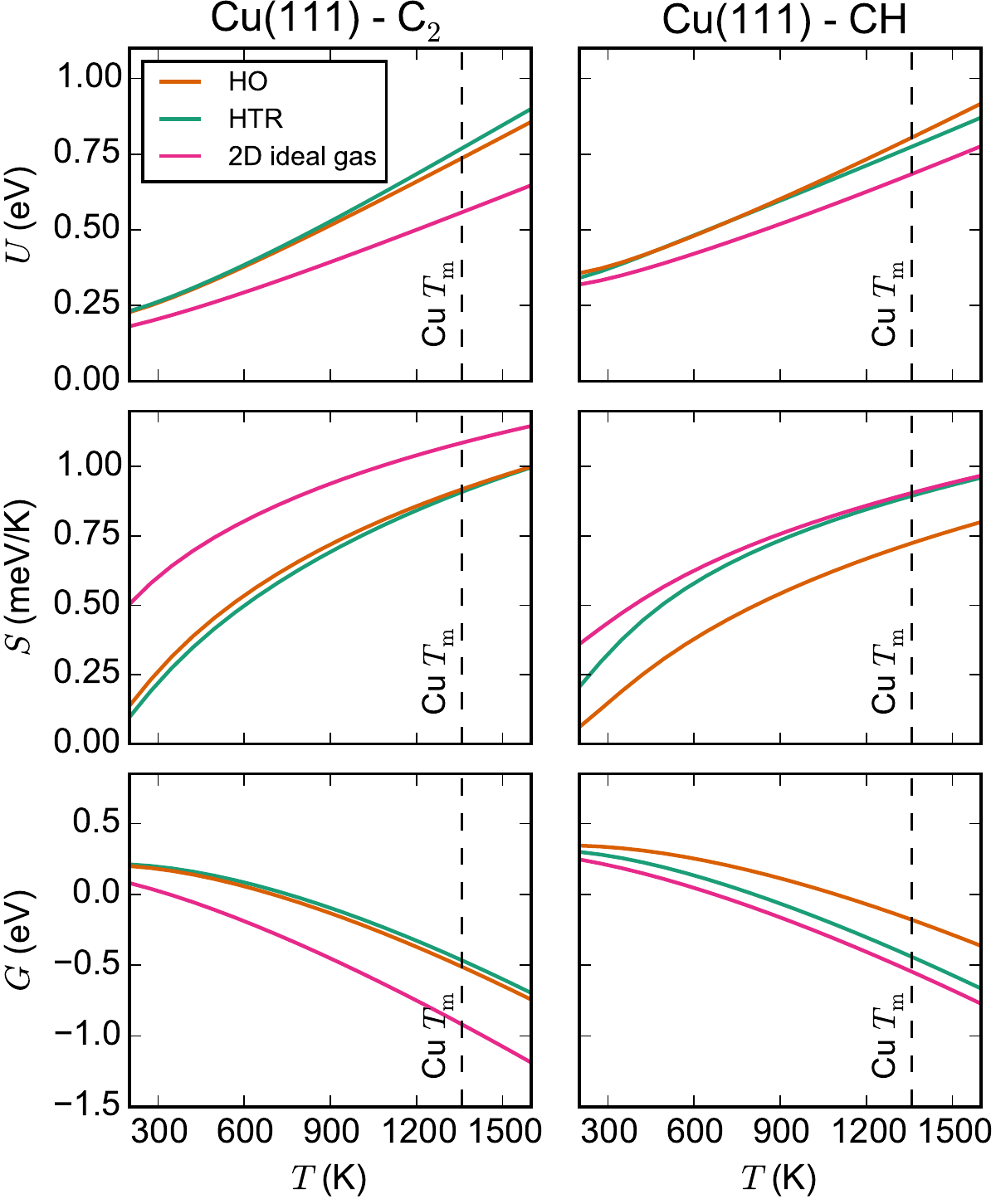}
\caption{Comparison of the internal energy $U$, the entropy $S$, and the Gibbs free energy $G$ in the harmonic oscillator (HO) model, the hindered translator / rotator (HTR) model, and the 2D ideal gas model for C$_2$ (left panels) and CH (right panels) at Cu(111). The Cu melting temperature ($T_{\rm m}$) of 1358 K is indicated with the black dashed line.}
\label{Fig:3}
\end{figure}

Next, we compare the models for the estimation of thermodynamic properties of adsorbates on Cu discussed above. Fig.\ \ref{Fig:3} compares the internal energy, entropy and Gibbs free energy of C$_2$ and CH at Cu(111) in the HO, HTR and 2D ideal gas models as a function of temperature. Following the ASE nomenclature, the internal energy $U$ is defined as
\begin{equation}\label{eq:5}
U(T) = E_{\rm pot} + E_{\rm ZPE} + E_{\rm vib}(T) + E_{\rm trans}(T) + E_{\rm rot}(T) \quad ,
\end{equation}
where $E_{\rm pot}$ is the DFT-calculated potential energy (set to zero in Fig.\ \ref{Fig:3}), $E_{\rm ZPE}$ contains all zero-point energy corrections from the vibrational as well as hindered translational and rotational modes, and the last three terms describe the increase in energy of the system due to population of excited vibrational, translational and rotational states at finite temperatures. Here the contributions from translational and rotational modes are only present in the HTR and 2D ideal gas models.
The entropy $S$ is simply the sum of the vibrational, translational and rotational contributions
\begin{equation}\label{eq:6}
S(T) = S_{\rm vib}(T) + S_{\rm trans}(T) + S_{\rm rot}(T) \quad ,
\end{equation}
and the Helmholtz free energy $F$ is given as
\begin{equation}\label{eq:7}
F(T) = U(T) - T S(T) \quad .
\end{equation}
We approximate the Gibbs free energy $G$, formally given as
\begin{equation}\label{eq:8}
G(T) = F(T) + pV \quad ,
\end{equation}
with the Helmholtz free energy, as in the differences of free energies considered here the small volume dependence of the $pV$ term largely cancels out.

%%%% FIGURE %%%% 
%%%% FIGURE %%%% 

CVD growth of graphene on both solid and liquid Cu is typically carried out at a temperature rather close to the melting temperature of Cu of 1358 K (solid Cu CVD typically at around 1300 K \cite{Li2009,Bhaviripudi2010,Vlassiouk2011,Kim2012,Li2016,Huet2017} and liquid Cu CVD typically at around 1370 K \cite{Geng2012,Wu2012,Wu2013,Geng2014,Zeng2016,Xue2019,Zheng2019}). For C$_2$ the diffusion and rotation barrier of 0.56\,eV is significantly larger than $k_{\rm B} T$ at the melting temperature of Cu (0.12 eV), while the CH diffusion barrier of 0.12 eV is of equal magnitude to $k_{\rm B} T$. Consequently, it is observed in Fig.\ \ref{Fig:3} that the thermodynamic properties of C$_2$ are described rather similarly in the HO and the HTR model, while for CH the Gibbs free energy is about 0.3\,eV lower in the HTR model than in the HO model at $T_{\rm m}$, mainly due to the increased entropy of CH in the HTR model. In fact, at this temperature the free energy of CH is only by about 0.1\,eV higher in the HTR model than in the 2D ideal gas model where the diffusion barrier goes to zero. It is important to note here that while the HTR model reduces exactly to the 2D ideal gas model as the barriers go to zero, the HTR and HO models are not guaranteed to agree in the limit of large barriers. The reason for this is that the HTR model assumes a sinusoidal shape of the potential energy surface near the minimum (see Fig.\ \ref{Fig:2}(b))\cite{Campbell2016} and takes into account anharmonic effects through this assumed shape, while the HO model directly probes the curvature of the potential energy surface near the minimum through the finite-difference scheme employed in the calculation of the vibrational frequencies, but neglects any anharmonic effects. In this situation, we choose to consistently employ the HTR model in the \textit{ab initio} thermodynamics study of hydrocarbons at solid Cu surfaces discussed in the next section, both for systems with low and high diffusion and rotation barriers. In the prior low-barrier case, the HTR model correctly reduces to the 2D ideal gas model limit. In the latter high-barrier case, the difference to the HO model in the limit of large barriers is insignificant for the conclusions put forward below.

\subsection{Solid Cu CVD growth}
Since the discovery of a low-pressure CVD (LPCVD) growth protocol for achieving large-area and high-quality graphene on solid Cu foils by the Ruoff group in 2009 \cite{Li2009}, this method arguably remains one of the most popular growth protocols used intensively by both academic groups and industry. Here, low pressure refers to the fact that the pressure in the reaction chamber is kept in the millitorr range (typically around 0.5 mbar). Methane is often used as the carbon source owing to its high thermal stability against pyrolysis in the gas phase, which is preferred to achieve surface-mediated growth self-limited to a single layer of graphene \cite{Li2016}. While graphene growth could in principle be realized from methane alone, hydrogen is typically also added to the reaction chamber during growth. However, the exact role of hydrogen for the growth remains an open question. The influence of quite diverse ratios of CH$_4$ to H$_2$ partial pressures ranging from about $10^{-3}$ to 10 has been explored in recent experimental studies \cite{Li2009,Bhaviripudi2010,Vlassiouk2011,Kim2012,Huet2017}, while on the theoretical side a number of recent studies have focused on explaining the role of hydrogen for the stability of hydrocarbon adsorbates during growth \cite{Zhang2011,Zhang2014,Shu2015,Li2017}. In all these previous theoretical studies internal energy and entropy corrections to adsorbed species were either ignored or treated in the HO model. This makes the accuracy at typical solid Cu CVD growth temperatures at best questionable for species with low diffusion or rotation barriers, where we have shown above that the HTR model provides a more accurate description.

\begin{figure*}
\centering
\includegraphics[width=0.8\textwidth]{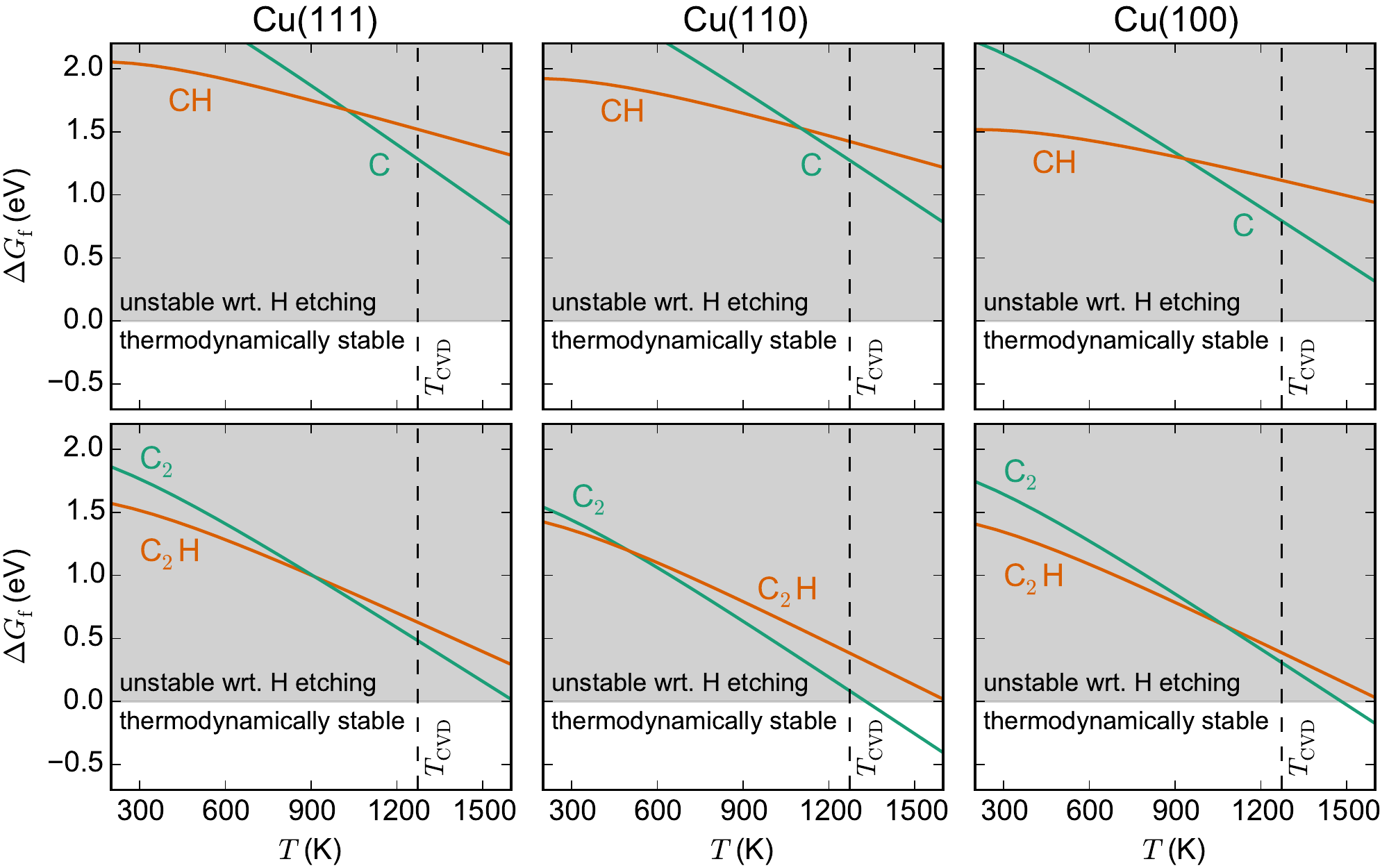}
\caption{Formation free energy ($\Delta G_{\rm f}$) evaluated in the HTR model as a function of temperature for hydrogenated C1 species (upper panels) and C2 species (lower panels) at the three low-index Cu facets. The gas phase pressures of CH$_4$ and H$_2$ are fixed to $1.3 \times 10^{-3}$ mbar and 0.47 mbar, respectively, which are low-pressure growth conditions used in Ref.\ \cite{Vlassiouk2011}. The region where the species are unstable with respect to hydrogen etching is shaded dark, and a typical solid Cu CVD growth temperature of 1273 K is indicated with the black dashed line.}
\label{Fig:4}
\end{figure*}

In Fig.\ \ref{Fig:4} we correspondingly explore the thermodynamic stability of various hydrogenated C1 and C2 species in the HTR model at the three considered facets Cu(111), Cu(110), and Cu(100) as a function of temperature for fixed low-pressure growth conditions at a CH$_4$ to H$_2$ ratio of $2.8 \times 10^{-3}$ as for instance employed in Ref.\ \cite{Vlassiouk2011}. From the definition of $\Delta G_{\rm f}$ (Eq.\ \ref{eq:1}) the species with the lowest Gibbs free energy is the most stable. Apart from the species explicitly shown in  Fig.\ \ref{Fig:4}, we have also carried out DFT caculations of the energies and vibrational frequencies for the C1 species CH$_2$ and CH$_3$ and the C2 species C$_2$H$_2$. However, at the growth temperature of 1273 K used in Ref.\ \cite{Vlassiouk2011}, these species are not among the most stable species at any of the facets irrespective of the diffusion and rotation barriers used in the HTR model. This also holds when varying the methane and hydrogen partial pressures within the ranges used in typical solid Cu CVD growth (to be discussed below). We therefore do not consider these species any further in this section. 

%%%% FIGURE %%%% 
%%%% FIGURE %%%% 

Considering now the stability trends of the different species plotted in Fig.\ \ref{Fig:4}, it is seen that the less hydrogen a species contains, the more it becomes stabilized at higher temperatures. This can be understood by considering that $\Delta G_{\rm f}$ (Eq.\ \ref{eq:1}) describes the Gibbs free energy change (per carbon atom) of the following reaction

\begin{scheme}
\ce{\textit{x} CH_{4}(g) <=> C_{\textit{x}}H_{\textit{y}}(ads) + (2 \textit{x} - \sfrac{\textit{y}}{2}) H_{2}(g)} \quad .
\caption{Formation of hydrocarbon adsorbate and gas-phase hydrogen from gas-phase methane (forward reaction) or hydrogen etching of hydrocarbon adsorbate to form gas-phase methane (reverse reaction).}
\label{Scheme1}
\end{scheme}

For the completely dehydrogenated species, the carbon monomer in the case of C1 species and the carbon dimer in the case of C2 species, $y$ is equal to zero and the reaction therefore consumes $x$ methane molecules and produces $2x$ hydrogen molecules. This net increase in the number of gas-phase species means that the entropy increases during the reaction, and the reaction therefore becomes more favorable (i.e.\ has a lower $\Delta G_{\rm f}$) at higher temperatures. The net increase of gas-phase molecules is smaller for the hydrogenated species CH and C$_2$H and their Gibbs free energies therefore decrease less with the temperature. At typical CVD growth temperatures this results in the completely dehydrogenated species being most stable, while the hydrogenated species become most stable only at several hundreds of Kelvin lower temperatures. Furthermore, it is seen that all species are thermodynamically unstable with respect to hydrogen etching (i.e.\ $\Delta G_{\rm f} > 0$). Depending on the kinetic barriers for the methane dehydrogenation and graphene growth steps, they might nevertheless be stable for a finite amount of time, which then allows for a certain proportion of these species to form and further react to form graphene flakes large enough to become thermodynamically stable (to be further discussed below).

\begin{figure*}
\centering
\includegraphics[width=0.8\textwidth]{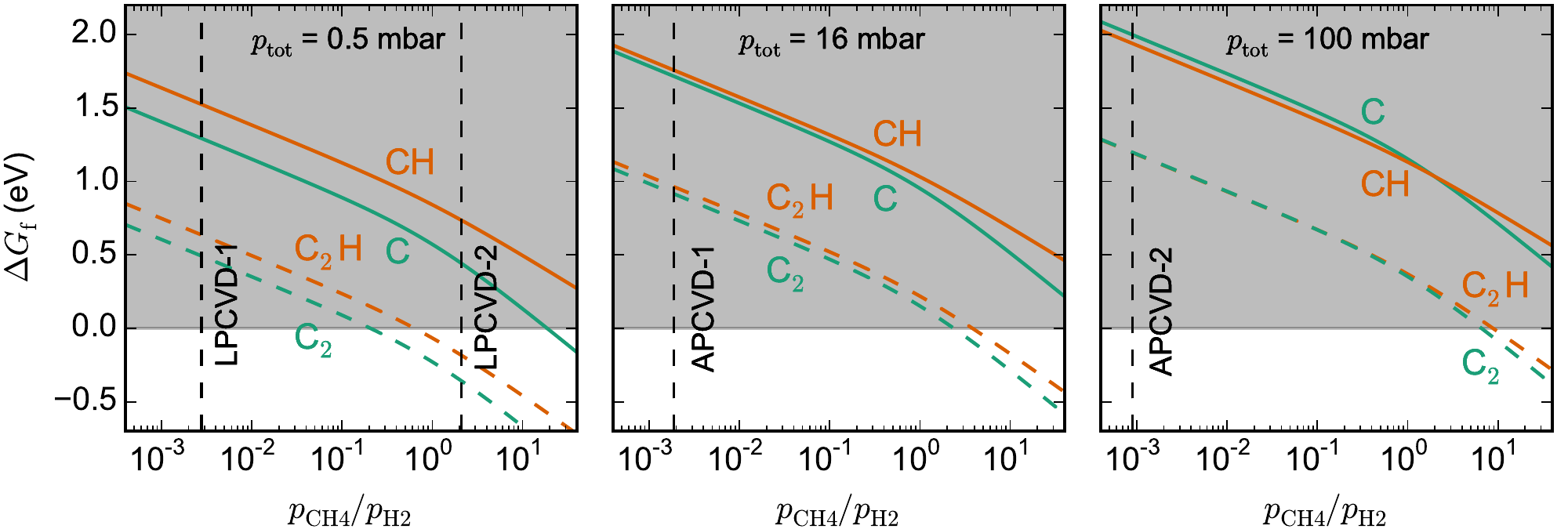}
\caption{Formation free energy ($\Delta G_{\rm f}$) evaluated in the HTR model as a function of the $p_{\rm CH_4}$ to $p_{\rm H_2}$ ratio for C, CH, C$_2$ and C$_2$H at Cu(111) and a temperature of 1273 K. The total ($p_{\rm CH_4}+p_{\rm H_2}$) pressure $p_{\rm tot}$ increases from the left to the right panels. Typical low-pressure and ambient-pressure CVD (LPCVD and APCVD) $p_{\rm CH_4}$ to $p_{\rm H_2}$ ratios from Ref.\ \cite{Vlassiouk2011} (LPCVD-1 and APCVD-1) and from Ref.\ \cite{Bhaviripudi2010} (LPCVD-2 and APCVD-2) are marked with black dashed lines. The region where the species are unstable with respect to hydrogen etching is shaded dark.}
\label{Fig:5}
\end{figure*}

To further explore the specific role of the methane and hydrogen partial pressures, we plot in Fig.\ \ref{Fig:5} the stability of C, CH, C$_2$ and C$_2$H at Cu(111) as a function of the CH$_4$ to H$_2$ ratio for three different total ($p_{\rm CH_4}+p_{\rm H_2}$) pressures and fixed temperature, chosen to allow direct comparison with existing experimental data \cite{Vlassiouk2011,Bhaviripudi2010}. It is seen that for either a decrease in the CH$_4$ to H$_2$ ratio or an increase in the total pressure, the hydrogenated species tend to become slightly more stable with respect to the dehydrogenated species. This can be understood by considering the Gibbs free energy change
(Eq.\ \ref{eq:1}) of the reaction in Scheme \ref{Scheme1}. The term $-\frac{y}{2 x} G_{\rm H_2}$ is non-zero only for the hydrogenated species and leads to a decrease of $\Delta G_{\rm f}$ as the Gibbs free energy of H$_2$(g) increases with the H$_2$ pressure as a result of either a lower CH$_4$ to H$_2$ ratio for fixed total pressure or an increase of the total pressure for fixed CH$_4$ to H$_2$ ratio. However, the effect of these pressure variations on the stability ordering of the hydrocarbons is in general much smaller than the effect on the overall stability of all hydrocarbon adsorbates, which is seen to strongly decrease for either a decrease in the CH$_4$ to H$_2$ ratio or an increase in the total pressure. The reason for this is that the term which depends on the degree of hydrogenation $y$, $-\frac{y}{2 x} G_{\rm H_2}$ in Eq.\ \ref{eq:1}, is of much smaller magnitude than the two terms which are independent of $y$, $+2 G_{\rm H_2}$ and $-G_{\rm CH_4}$ in Eq.\ \ref{eq:1}. For a decrease in the CH$_4$ to H$_2$ ratio the latter two terms both become more positive, giving rise to a strong increase in $\Delta G_{\rm f}$. For an increase of the total pressure, the first hydrogen-related term becomes more positive and the second methane-related term becomes more negative. Overall this leads to an increase in $\Delta G_{\rm f}$ since the change in the hydrogen-related term, which involves two molecules, is of larger magnitude than the change in the methane-related term, which involves only one molecule.

%%%% FIGURE %%%% 
%%%% FIGURE %%%% 

The lowest total pressure investigated in Fig.\ \ref{Fig:5}(a) (0.5 mbar) corresponds to typical LPCVD growth conditions. Two different CH$_4$ to H$_2$ ratios used in literature studies \cite{Bhaviripudi2010,Vlassiouk2011}, both resulting in single-layer graphene flakes, are highlighted. For the lower CH$_4$ to H$_2$ ratio used in Ref.\ \cite{Vlassiouk2011} it was observed that higher methane proportions resulted in multi-layer growth, while this was apparently not the case in Ref.\ \cite{Bhaviripudi2010}. This highlights the difficulty in comparing growth results from different experimental setups, where many uncontrolled factors might influence the graphene growth. For example, a well-known role of hydrogen is to keep the Cu surface free from oxidation arising from oxygen or water contamination in the reaction chamber \cite{Li2016}. A higher leak rate in a given experiment might thus necessitate a higher concentration of hydrogen for the growth. 

Nevertheless, the increased tendency to form multi-layer graphene flakes at higher CH$_4$ to H$_2$ ratios could be related to the higher stability of all adsorbates at these conditions. If assuming first-order kinetics and a simple Langmuir adsorption model \cite{Chorkendorff} (i.e.\ assuming at most one adsorbate per site and ignoring all interactions between the adsorbates) the coverage $\Theta_A$ of a hydrocarbon species $A$ can be calculated as
\begin{equation}\label{eq:9}
\Theta_A = \frac{K_A}{1+K_A} \quad ,
\end{equation}
where the dependence on the gas-phase pressures is incorporated into the equilibrium constant $K_A$ given as
\begin{equation}\label{eq:10}
K_A = \exp \left( \frac{-\Delta G_{{\rm f},A}}{k_{\rm B} T} \right) \quad .
\end{equation}
For a large positive $\Delta G_{\rm f}$ (low CH$_4$ to H$_2$ ratio) the coverage will thus be much smaller than one, while for a large negative $\Delta G_{\rm f}$ (high CH$_4$ to H$_2$ ratio) the coverage will be close to one. While the kinetic barriers of the ongoing chemical reactions, in particular graphene growth, and adsorbate-adsorbate interactions will in practice also affect the actual surface coverages, it can thus generally be expected that a lower $\Delta G_{\rm f}$ for a given species will lead to a higher surface coverage of that species. In this respect, a higher coverage of carbon species could lead to increased carbon dissolution into the bulk Cu, despite the nominal low carbon solubility of Cu. This, as well as faster kinetics of graphene growth, could all facilitate the nucleation of a second graphene layer below the first. It is also well-known that a higher CH$_4$ to H$_2$ ratio results in an increased graphene nucleation site density, which leads to overall smaller graphene domains and increasing amounts of structural defects arising from graphene grain boundaries in the final graphene sheet \cite{Huet2017}. This increased graphene nucleation rate could again be a result of a higher coverage of carbon species at higher CH$_4$ to H$_2$ ratios.

The higher total pressures considered in Fig.\ \ref{Fig:5}(b) and (c) (16 and 100 mbar, respectively) correspond to ambient-pressure CVD (APCVD) studies, also from Refs.\ \cite{Bhaviripudi2010,Vlassiouk2011}, where Ar is used as a carrier gas. As already discussed, these higher total pressures are predicted to decrease the stability of all hydrocarbon adsorbates compared to LPCVD conditions at the same fixed CH$_4$ to H$_2$ ratio. Both APCVD studies employed rather low CH$_4$ to H$_2$ ratios and reported that higher ratios resulted in multi-layer growth. As for the LPCVD conditions, this could be explained by the increase in stability of all species at higher CH$_4$ to H$_2$ ratios. In general, the growth results of the two experimental studies considered here \cite{Bhaviripudi2010,Vlassiouk2011} suggest that LPCVD conditions are more robust against multi-layer growth, allowing for the use of a wider range of CH$_4$ to H$_2$ ratios compared to APCVD conditions. This is probably one of the reasons for the popularity of the LPCVD method \cite{Li2016}. The reason for the increased robustness of the LPCVD method is not clear from the theoretical results presented here and might arise from other factors not considered. For example, it has been suggested that mass transport within the reaction chamber, i.e.\ diffusion of gas-phase species through a boundary layer close to the surface, is the rate-limiting step under APCVD conditions, while for LPCVD instead the surface reactions are rate-limiting \cite{Bhaviripudi2010}. This means that geometric effects of the gas flow and of the geometry of the reaction chamber \cite{Matera2010,Matera2014} could play a much larger role under APCVD studies and significantly affect the actual pressures of methane and hydrogen in the vicinity of the surface, complicating the comparison to theoretical studies that only take into account the surface reactions. Possible other explanations include the cleanliness and density of defects of the employed Cu foils under LPCVD and APCVD conditions, the amount of oxidizing impurities in the gas phase etc.

Overall, our results show that the dehydrogenated species, the carbon monomer and dimer, are the most stable surface species at typical solid Cu CVD growth conditions. A possible exception is at the highest considered total pressures (APCVD conditions), where the difference in stability between hydrogenated and dehydrogenated species is lower than the expected accuracy of the theoretical method. Our results are thereby in disagreement with the theoretical studies from Ref.\ \cite{Li2017}, where it was found that the coverage of CH becomes higher than that of C (and even higher than that of C$_2$) at 1300 K, a total pressure of 13 mbar, and a CH$_4$ to H$_2$ ratio below 100. Most likely, this discrepancy is due to the complete negligence of internal energy and entropy corrections to the adsorbed species in Ref.\ \cite{Li2017}.

\subsection{Liquid Cu CVD growth}
In 2012 it was discovered that graphene CVD growth at a liquid Cu surface leads to large single-crystalline and single-layered graphene flakes of very high structural quality \cite{Geng2012,Wu2012}. The large size and single-crystalline nature of the flakes is a result of a drastically lowered graphene nucleation rate when the temperature is raised above the melting temperature of Cu \cite{Zheng2019}, which presumably arises from the smooth homogeneous nature of the liquid metal surface, as compared to the many defects and grain boundaries observed in the Cu foils and thin films used for solid Cu CVD. Graphene growth at liquid Cu surfaces has also been further investigated in a number of more recent studies \cite{Wu2013,Geng2014,Zeng2016,Xue2019}. Compared to the wide variety of growth conditions used in solid Cu CVD, rather similar ambient-pressure growth conditions were used in all of these liquid Cu CVD studies to date with temperatures ranging from the melting temperature of Cu (1358 K) to 1433 K, total ($p_{\rm CH_4}+p_{\rm H_2}$) pressures of about 100$-$1000 mbar (with or without Ar as a carrier gas), and CH$_4$ to H$_2$ ratios ranging from 0.001 to 0.02.

\begin{figure}
\centering
\includegraphics[width=\columnwidth]{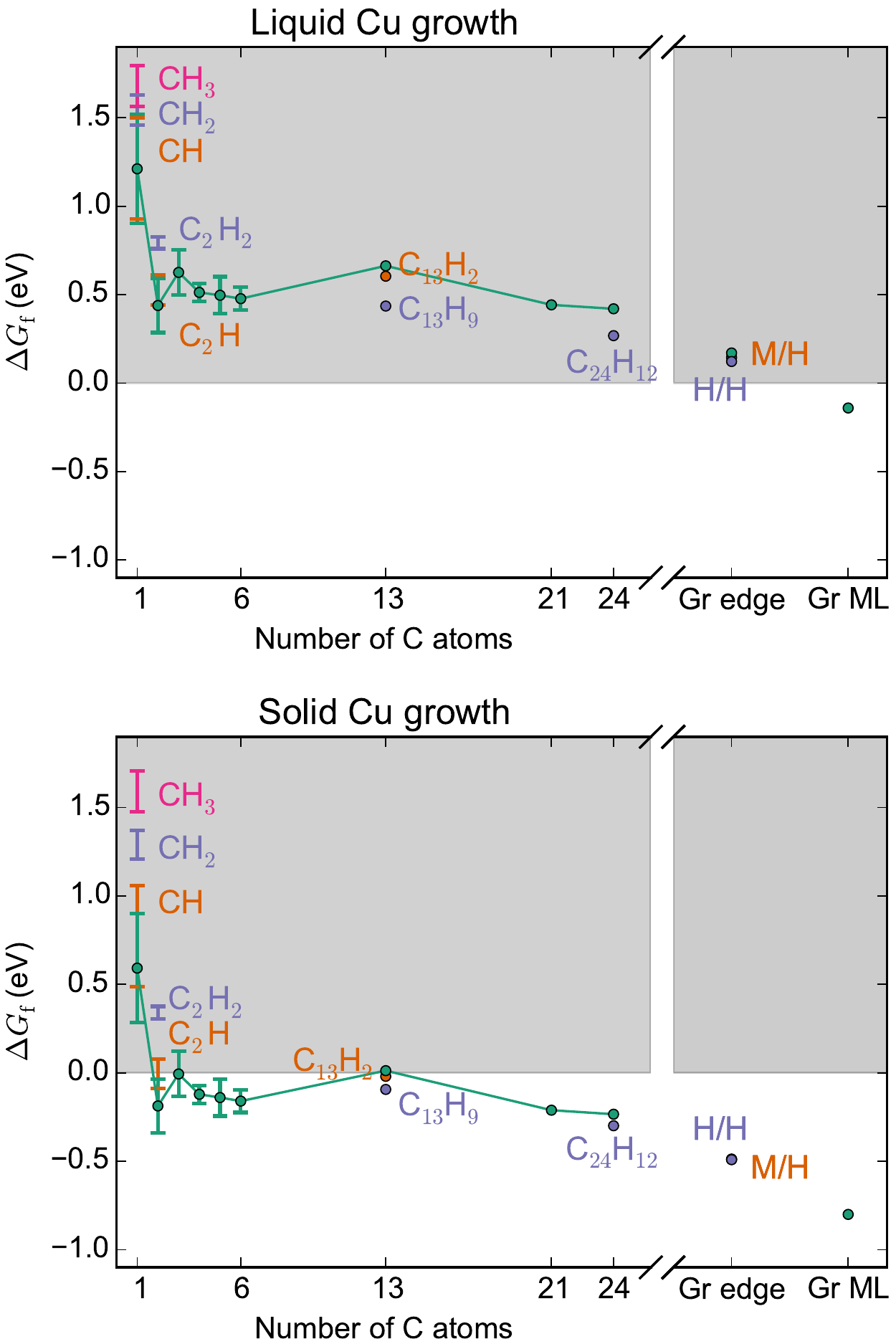}
\caption{Formation free energy ($\Delta G_{\rm f}$) for different sizes of hydrocarbon clusters as well as graphene (Gr) edges and the complete Gr monolayer (ML). In the upper panel typical liquid Cu CVD growth conditions ($T=1373$ K, $p_{\rm CH_4}=3.3$ mbar, $p_{\rm H_2}=330$ mbar) from Ref.\ \cite{Zeng2016} are used, while in the lower panel typical solid Cu LPCVD growth conditions ($T=1273$ K, $p_{\rm CH_4}=0.37$ mbar, $p_{\rm H_2}=3.73$ mbar) from Ref.\ \cite{Kim2012} are used. All hydrocarbon clusters are evaluated in the 2D ideal gas model, while Gr edges and the Gr ML are evaluated in the HO model as they are assumed to remain immobile on the surface. For the hydrocarbon clusters containing 1-6 C atoms the error bars represent the spread in the results obtained by using the formation energies and frequencies for the three different Cu facets, while for the larger clusters, the Gr edges and the Gr ML data from the Cu(111) facet is used. The solid green lines connect all pure carbon clusters. The region where the species are unstable with respect to hydrogen etching is shaded dark.}
\label{Fig:8}
\end{figure}

\begin{figure*}
\centering
\includegraphics[width=\textwidth]{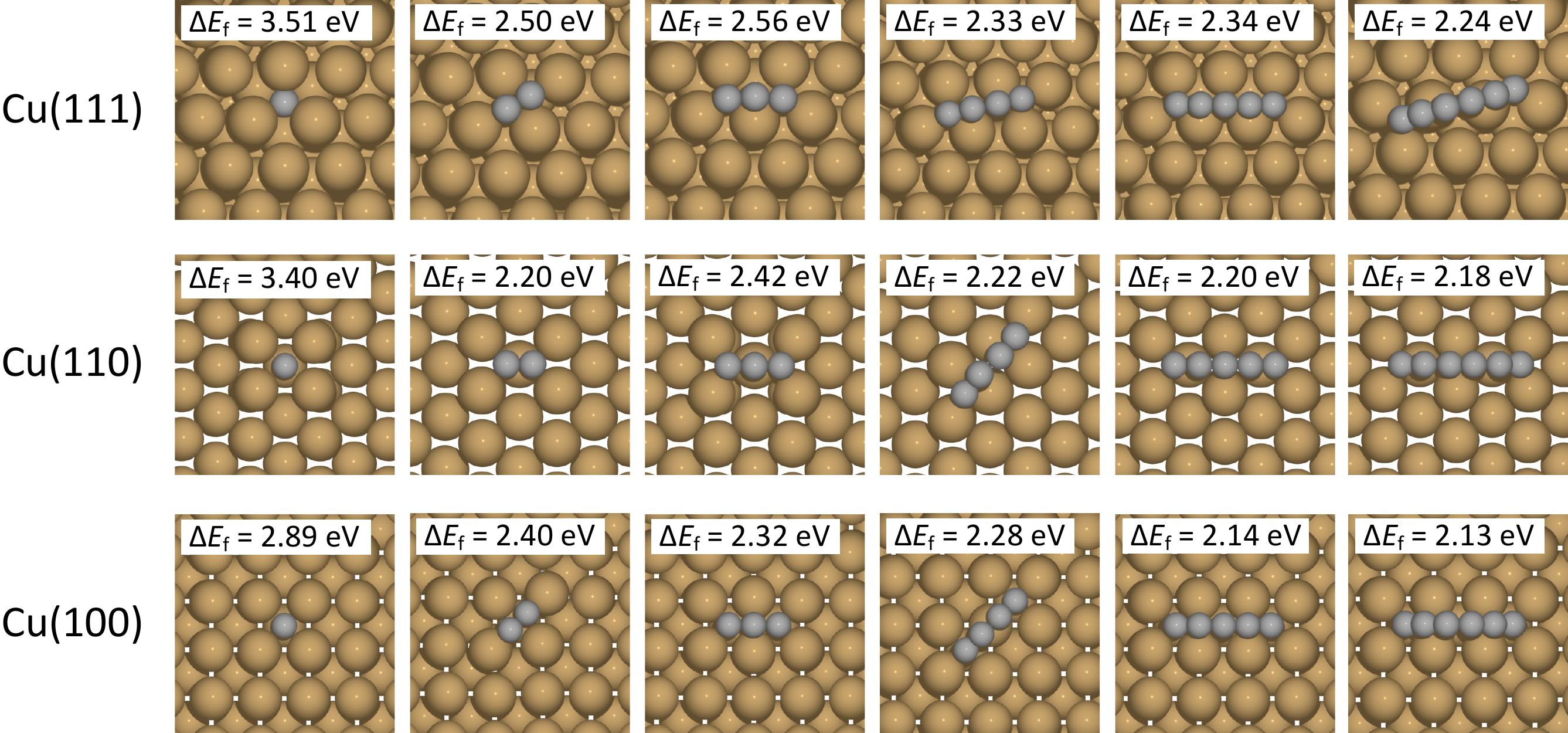}
\caption{Structure (top view) and formation energy ($\Delta E_{\rm f}$) of C$_1$-C$_6$ species at the three low-index Cu facets.}
\label{Fig:6}
\end{figure*}

\begin{figure}
\centering
\includegraphics[width=\columnwidth]{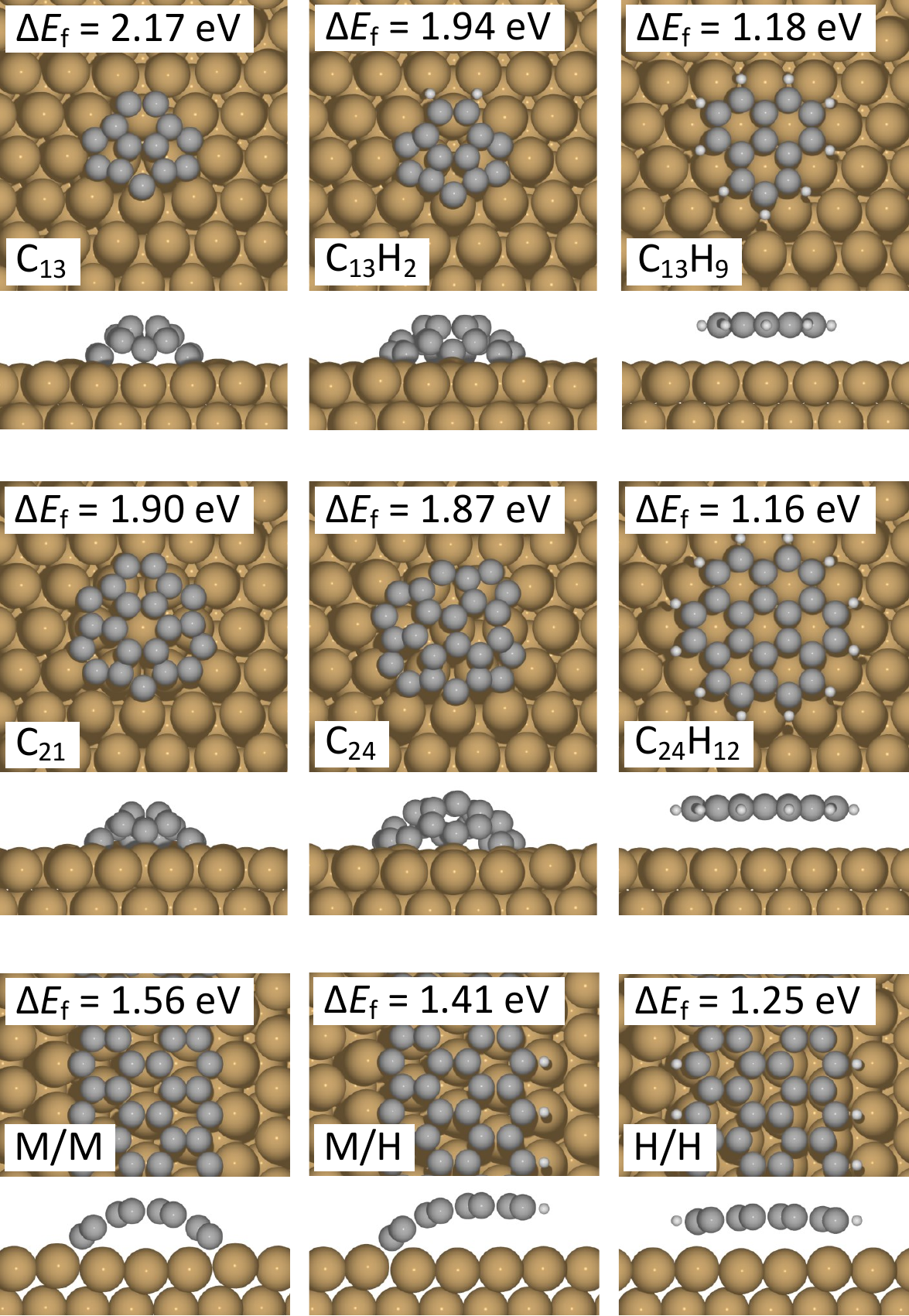}
\caption{Structure (top and side view) and formation energy ($\Delta E_{\rm f}$) of larger hydrocarbon clusters as well as metal (M)- and hydrogen (H)-terminated graphene edges at Cu(111).}
\label{Fig:7}
\end{figure}

In Fig.\ \ref{Fig:8} we investigate the stability of various sizes of hydrocarbon clusters all the way up to the full graphene monolayer (ML) for typical liquid Cu CVD growth conditions and compare to typical solid Cu LPCVD growth conditions. On a liquid Cu surface all hydrocarbon adsorbates are expected to be highly mobile and we therefore employ the 2D ideal gas model for internal energy and entropy corrections\footnote{Note that for ease of comparison we here use the 2D ideal gas model for all hydrocarbon clusters and reaction conditions, which artificially stabilizes especially smaller hydrocarbon clusters with large diffusion / rotation barriers at the solid Cu LPCVD conditions. For e.g.\ Cu(111) the model is expected to be a good approximation for weakly bound species such as CH$_3$, CH$_2$ and CH, but is less accurate for the more strongly bound species such as C and C$_2$.\label{footnote_1}}. The exact structure of the liquid Cu surface is not accessible from static DFT calculations. Instead, we assume that the three investigated solid Cu facets, Cu(111), Cu(110) and Cu(100), provide active site motifs that are representative for the distribution of active site motifs found on a liquid Cu surface. In addition to the C1 and C2 species discussed in the previous section, we here include also CH$_2$, CH$_3$ and C$_2$H$_2$ as well as the pure carbon clusters C$_3$-C$_6$ at all three facets (see structures and formation energies in Fig.\ \ref{Fig:6}). As also observed in a previous study \cite{Mi2012}, the formation energies become more and more facet-independent the larger the cluster. In the following we therefore focus the discussion on the Cu(111) facet for the larger C13, C21, C24 species, graphene edges and the full graphene ML (see structures and formation energies in Fig.\ \ref{Fig:7}), with analog conclusions to be drawn from the other two facets.

%%%% FIGURE %%%% 
%%%% FIGURE %%%% 

%%%% FIGURE %%%% 
%%%% FIGURE %%%% 

%%%% FIGURE %%%% 
%%%% FIGURE %%%% 

For the solid Cu growth conditions employed in Fig.\ \ref{Fig:8}, the graphene growth rate was shown to follow an Arrhenius-like behavior in the temperature range of 1023$-$1273 K with an apparent activation energy of $2.6 \pm 0.5$\,eV \cite{Kim2012}. Dissociative CH$_4$ adsorption was ruled out as the rate-limiting step based solely on the DFT-calculated barriers for the dehydrogenation steps at Cu(111), where the highest barrier of {\raise.17ex\hbox{$\scriptstyle\sim$}}2\,eV is encountered for CH dissociation \cite{Zhang2011}. However, taking into account also the thermodynamics of these reaction steps as done in Fig.\ \ref{Fig:8}, it can be seen that the species involved in CH$_4$ dissociation already have a significant positive Gibbs free energy of formation without even considering the kinetic barriers. Adding the DFT-calculated barriers for all dehydrogenation steps on Cu(111) from Ref.\ \cite{Zhang2011} to our calculated $\Delta G_{\rm f}$ values for the species on the Cu(111) facet, thereby assuming that internal energy and entropy corrections to the adsorbate and the TS are of similar magnitude, we find that the TS with the maximum free energy is the TS for CH dissociation (3.2\,eV) at 1023 K and the TS for CH$_3$ dissociation (also 3.2\,eV) at 1273 K. Given that the experimentally used Cu foil is polycrystalline and might contain defect sites with lower barriers for the CH$_4$ dissociation steps than those calculated for Cu(111), this theoretical activation energy is in reasonable agreement with the measured value. While we cannot rule out for certain that other reaction steps involved in the later growth processes might also have high barriers, the fact that all of the larger hydrocarbons have much lower and rather similar $\Delta G_{\rm f}$ values strongly suggests that methane adsorption and dehydrogenation is indeed the rate-limiting step for graphene growth. Consequently, the graphene growth rate is predicted to be highly sensitive to reaction parameters such as the CH$_4$ to H$_2$ ratio. As discussed previously, all hydrocarbon adsorbates become more stable at higher CH$_4$ to H$_2$ ratios, which is therefore predicted to lower the apparent activation energy and increase the growth rate. This is consistent with the experimental observation that the growth rate can be accelerated by increasing the methane concentration in the gas-phase mixture in both solid Cu CVD \cite{Vlassiouk2011} and liquid Cu CVD \cite{Zheng2019}. The findings discussed here for solid Cu CVD likely also hold for liquid Cu CVD, since the comparison in Fig.\ \ref{Fig:8} shows that the free energy profile is qualitatively similar in the two cases. 

As was also observed when comparing solid Cu LPCVD and APCVD growth conditions, it is seen in Fig.\ \ref{Fig:8} that all hydrocarbon species have an overall lower stability at the liquid Cu growth conditions, for which even higher total pressures than in solid Cu APCVD are typically used together with low CH$_4$ to H$_2$ ratios. As already discussed for APCVD, the higher pressures used for liquid Cu CVD might complicate the comparison to experimental results, since mass-transfer limitations in the gas phase could significantly affect the actual pressures of methane and hydrogen in the vicinity of the surface. Nevertheless, it is interesting to note that our results for typical liquid Cu CVD conditions indicate that graphene growth is carried out very close to the thermodynamic limit where the full graphene ML becomes unstable with respect to hydrogen etching, i.e.\ only very large graphene flakes will be thermodynamically stable. For solid Cu LPCVD conditions, on the other hand, already the carbon dimer might be thermodynamically stable (depending on the exact LPCVD conditions used, as discussed in the previous section\footref{footnote_1}). In this regard, it can be speculated that the proximity to the thermodynamic limit could result in a continuous decomposition of any defected carbon structures formed during the growth, since such structures might be thermodynamically unstable. This would improve the overall structural quality of the flake, consistent with the observed high structural quality of flakes grown at liquid Cu \cite{Geng2012,Wu2012}. Such defect-healing mechanisms might only be possible for Cu surfaces in the liquid state due to the high temperatures employed and the dynamic nature of the liquid, which might serve to lower the kinetic barriers for such defected structures to decompose \cite{Li2014}. Thus, it might not be possible under solid Cu APCVD conditions, even if our analysis of these growth conditions in the previous section suggested a similar low stability of all hydrocarbon adsorbates.
 
Another interesting point to note in Fig.\ \ref{Fig:8} is that the carbon dimer represents a local minimum on the free energy curve, i.e.\ it is predicted to be present at the surface in higher coverages than the other smaller carbon clusters. While we consider here only the thermodynamics, literature DFT calculations on Cu(111) have shown that when two carbon monomers are placed into neighboring adsorption sites, C$_2$ formation proceeds barrier-less for most co-adsorption configurations \cite{Riikonen2012}. The carbon dimer is therefore a good candidate feeding species for the continued growth of already nculeated graphene flakes, as has also been suggested based on calculations of barriers for edge-attachment processes and kinetic Monte Carlo simulations \cite{Wu2015}.

In Fig.\ \ref{Fig:8} we plot also the stability of a number of hydrogenated clusters as well as hydrogen-terminated graphene edges. At the liquid Cu surface the C and CH clusters are of similar stability, and they might therefore both be formed. The C$_2$ cluster is however more stable than the C$_2$H cluster (averaged over the three active site motifs considered). For the larger clusters the complete hydrogenation of all edge carbon atoms, i.e.\ the formation of the C$_{13}$H$_{9}$ and C$_{24}$H$_{12}$ clusters, is thermodynamically favorable. For solid Cu LPCVD conditions the completely dehydrogenated C1 and C2 species are most stable, as already noted in the previous section, whereas the hydrogenated larger clusters are slightly more stable than the dehydrogenated clusters. For both solid and liquid Cu CVD growth conditions the stability of the graphene edges is the same or very similar regardless of the degree of hydrogenation. Several theoretical studies from the literature have proposed that hydrogen-terminated edges should be most stable at low CH$_4$ to H$_2$ ratios \cite{Zhang2014,Li2017}, however, based on our results this conclusion cannot be definitively supported. The similar stability of all edge terminations could in practice mean that the formation of both metal-terminated and hydrogen-terminated clusters is possible during graphene growth both under liquid and solid Cu CVD growth conditions. However, the clusters with hydrogen-terminated edges are not necessarily stable against desorption. In Table \ref{tab:3} we give the DFT adsorption energies of various hydrocarbon clusters at the Cu(111) facet as well as the calculated lifetime (inverse rate of desorption, see Eq.\ \ref{eq:4}) at a typical liquid Cu CVD growth temperature of 1370 K. The adsorption energies of C$_{13}$H$_{9}$ and C$_{24}$H$_{12}$ are very low compared to the dehydrogenated clusters, since they are only bound to the surface by weak vdW interactions. Consequently, the calculated lifetime is very short (in the nano- to microsecond range). Hence, depending on the relevant timescales for the graphene growth processes, such clusters might desorb before they are able to grow large enough to become stable. The small CH and C$_2$H clusters, on the other hand, have lifetimes in the range of seconds to hours, and they are therefore more likely to play a role for graphene growth. 

%%%% TABLE %%%% 
\begin{table}
  \centering
  \begin{tabular}{lll}
	  \hline\hline
		Species & $\Delta E_{\rm ads}$ & $\tau$ \\
		\hline
		C & -5.51 & $5.6 \times 10^6$ \\
		CH & -5.23 & $2.7 \times 10^4$ \\
		CH$_2$ & -3.38 & $7.2 \times 10^{-3}$ \\
		CH$_3$ & -1.78 & $1.4 \times 10^{-8}$ \\
		C$_2$ & -6.36 & $2.1 \times 10^{9}$ \\
		C$_2$H & -4.33 & 3.5 \\
		C$_2$H$_2$ & -1.65 & $3.1 \times 10^{-8}$ \\
		C$_{13}$H$_9$ & -1.39 & $5.8 \times 10^{-9}$ \\
		C$_{24}$H$_{12}$ & -2.01 & $1.2 \times 10^{-7}$ \\
		\hline\hline
  \end{tabular}
  \caption{Adsorption energy ($\Delta E_{\rm ads}$ in eV) and lifetime ($\tau$ in s) evaluated in the 2D ideal gas model at 1370 K for various hydrocarbons at Cu(111).}
  \label{tab:3}
\end{table}
%%%% TABLE %%%% 

A further argument against the presence of larger hydrogenated clusters at liquid Cu CVD growth conditions is the fact that the structure of the dehydrogenated clusters calculated for the Cu(111) facet might not be the most stable structure at the liquid Cu surface. At solid Cu, the most stable pure carbon clusters contain 5-membered rings, which allows for the unsaturated edge carbon atoms to bend down towards the surface and form covalent bonds to the surface Cu atoms (see Fig.\ \ref{Fig:7}). For a liquid Cu surface, on the other hand, the dynamic state of the Cu atoms might allow for flat defect-free clusters that achieve Cu-coordination of the edge C atoms by slightly sinking down into the liquid \cite{Cingolani2019}. Such defect-free geometries could possibly stabilize the pure carbon clusters to the extent where their stability would become similar or even greater than the clusters with hydrogenated edges. Less defects in the pure carbon clusters formed during graphene growth on liquid Cu, as compared to on solid Cu, could also be one of the factors explaining the higher structural quality of graphene flakes grown on liquid Cu.

\section{Conclusions}
In this work we employed \textit{ab initio} thermodynamics together with literature models for the estimation of thermodynamic properties of adsorbates, in particular the HO, HTR and 2D ideal gas models, to explore the role of various hydrocarbon adsorbates in graphene growth on liquid and solid Cu surfaces for a wide range of experimentally used CVD growth conditions. We find that thermodynamic considerations in general, and in particular the choice of an appropriate thermodynamic model, is highly important for explaining experimental findings. We considered a wide range of hydrocarbon adsorbates and explored variations in their thermodynamic stability with typical growth parameters such as temperature, CH$_4$ to H$_2$ ratio and total ($p_{\rm CH_4}+p_{\rm H_2}$) pressure. Experimental observations such as an increased graphene nucleation rate and susceptibility to multi-layer growth in studies employing a high CH$_4$ to H$_2$ ratio were rationalized based on the theoretically predicted higher thermodynamic stability of all hydrocarbon adsorbates at these conditions.

For typical solid Cu LPCVD growth conditions our results showed that the completely dehydrogenated C1 and C2 species, i.e.\ the carbon monomer and dimer, are the most stable species. However, smaller hydrogenated species such as CH and C$_2$H could also play a role for large total pressures such as employed in solid and liquid Cu APCVD growth. The carbon dimer was shown to have a significantly higher stability than the other smaller hydrocarbon clusters both for solid and liquid CVD growth conditions, and could thus be the feeding species responsible for the continued growth of already nculeated graphene flakes. For such medium- to large-sized graphene flakes our results cannot definitively rule out that the edges could be terminated with hydrogen under both solid and liquid Cu growth conditions. However, formed hydrogen-terminated clusters might desorb from the Cu surface before growing large enough to become thermodynamically stable, since they are only bound to the surface by weak van der Waals interactions.

Finally, we constructed a free energy diagram of hydrocarbon species all the way from the reaction intermediates involved in methane dehydrogenation to the full graphene ML. Based on this diagram as well as literature calculations of methane dehydrogenation barriers, we suggest that methane adsorption and dehydrogenation is the rate-limiting step for graphene growth, giving rise to a theoretical activation energy consistent with experimental measurements. This step is expected to be rate-limiting both for solid and liquid Cu CVD growth, and the growth rate is predicted to be highly dependent on the growth parameters such as in particular the CH$_4$ to H$_2$ ratio, which is also consistent with experimental findings. Regarding liquid Cu CVD under typical growth conditions, the high structural quality of grown graphene samples is suggested to be related to the fact that the growth is typically carried out very close to the thermodynamical limit where the full graphene ML becomes unstable to hydrogen etching, defect-healing mechanisms made possible by the liquid state of the surface \cite{Li2014}, as well as the possibility that carbon clusters with defects (e.g.\ 5-membered ring defects) might be less likely to form on a liquid Cu compared to a solid Cu surface.

\begin{suppinfo}
Supporting Information. Energies, vibrational frequencies, structures and graphene phonon density of states. This material is available free of charge via the Internet at http://pubs.acs.org.
\end{suppinfo}

\begin{acknowledgement}
This project has received funding from the European Union’s Horizon 2020 research and innovation programme under grant agreement 736299. Responsibility for the information and views set out in this article lies entirely with the authors. The authors gratefully acknowledge the Gauss Centre for Supercomputing e.V. (www.gauss-centre.eu) for funding this project by providing computing time on the GCS Supercomputer SuperMUC at Leibniz Supercomputing Centre (www.lrz.de) as well as through the John von Neumann Institute for Computing (NIC) on the GCS Supercomputer JUWELS \cite{JUWELS} at J{\"u}lich Supercomputing Centre (JSC).
\end{acknowledgement}

\bibliography{AITD_references}

\end{document}